\documentstyle[12pt]{article}
\newcommand{\be}{\nopagebreak\begin{equation}}
\newcommand{\ee}{\end{equation}}
\newcommand{\ba}{\begin{array}}
\newcommand{\ea}{\end{array}}
\newcommand{\bp}{\begin{picture}}
\newcommand{\ep}{\end{picture}}
\newcommand{\eol}{\\[1pc]}
\newcommand{\rf}[1]{Ref.~\cite{#1}}
\newcommand{\eq}[1]{Eq.~(\ref{#1})}
\newcommand{\bi}[6]{\bibitem{#1}{#2, }{\sl #3 }{\bf #4}{ (#5)}{ #6}}
\newcommand{\wt}[1]{\widetilde{#1}}

\newcommand{\ie}{{\it i.e., }}
\newcommand{\eg}{{\it e.g., }}
\renewcommand{\d}{\partial}
\newcommand{\tr}{{\rm tr}}
\newcommand{\scr}{\scriptstyle}
\newcommand{\scs}{\scriptscriptstyle}
\newcommand{\mv}[1]{\langle #1 \rangle}

\begin{document}
\begin{titlepage}
\begin{flushright}
NBI-HE-94-37\\
August, 1994\\
\end{flushright}
\vspace*{36pt}
\begin{center}
{\huge \bf  On entropy of 3-dimensional simplicial complexes}
\end{center}
\vspace{2pc}
\begin{center}
 {\Large D.V. Boulatov}\\
\vspace{1pc}
{\em The Niels Bohr Institute,
University of Copenhagen\\
Blegdamsvej 17, DK-2100 Copenhagen \O,
Denmark\\
and\\
International School for Advanced Studies (SISSA/ISAS)\\
via Beirut 2-4, I-34014 Trieste, Italy.\footnote{$\mbox{}^)$
Present address.}}
\vspace{2pc}
\end{center}
\begin{center}
{\large\bf Abstract}
\end{center}
We prove that the number of 3-dimensional simplicial complexes having
the spherical topology grows exponentially as a function of a volume.
It is suggested that the 3d simplicial quantum gravity has
qualitatively  the same phase structure  as the O(n) matrix model in
the dense phase.
\vfill
\end{titlepage}

\section{Introduction}

The success of the matrix models as a theory of 2D quantum gravity and
non-critical strings has brought about a hope that analogous discrete
approaches might be instructive in higher dimensions as well. The most
natural way to introduce discretized quantum gravity is to consider
simplicial complexes instead of continuous manifolds. Then, a path
integral over metrics (which is the crux of any approach to quantum
gravity) can be simply defined as a sum over all complexes having some
fixed properties. For example, it is natural to restrict their topology.
In the present paper, we consider only the 3-dimensional case, where,
by definition, a simplicial complex is a collection of tetrahedra glued
by their faces in such a way that any two of them can have at most one
triangle in common.

To introduce metric properties, one can assume that complexes
represent piece-wise linear manifolds glued of equilateral tetrahedra
\cite{Regge}.
The volume is proportional to the number of them.
In a continuum limit, as this number grows, the length of links, $a$,
simultaneously tends to zero: $a\to0$. In three dimensions, there is
the one-to-one correspondence between topological classes of piece-wise
linear and smooth manifolds, therefore the model is
self-consistent.

Within a piece-wise linear approximation, the curvature is singular: the
space is flat everywhere off links. Therefore, one should consider
integrated quantities, the main of which is the mean scalar curvature

\be
\int d^3x\; \sqrt g R = a\Big(2\pi N_1 -6N_3\arccos \frac13\Big)
\ee
$a$ is the lattice spacing; $N_0,\ N_1,\ N_2,\ N_3$ are the numbers of
vertices, links, triangles and tetrahedra,  respectively.

For manifolds in 3 dimensions, the Euler character vanishes
$\chi = N_0-N_1+N_2-N_3=0$.
Together with the other constraint $N_2=2N_3$, it means that only 2 of
$N_i$'s are independent, $N_1=N_0+N_3$, and a natural lattice action
depends on 2 dimensionless parameters. Now, we are in a position to
define the partition function for simplicial gravity
\cite{ADJ,3dsim,3dsim2,B}:

\be
{\cal Z}(\alpha,\mu)=\sum_{\{S^3\}} e^{\alpha N_0 - \mu N_3}
=\sum_{N_3}^{\infty} Z_{N_3}(\alpha) e^{-\mu N_3}
\label{Z}
\ee
where, for concreteness, we have restricted the topology of complexes:
$\sum_{\{S^3\}}$ denotes the sum over all simplicial 3-spheres, $S^3$.

It can be shown that, for $N_3$ fixed, $N_0$ is restricted from above as
$N_0<\frac13N_3+const$. Therefore, it is natural, keeping $\alpha$ fixed, to
tend $\mu$ to its critical value, $\mu_c$, when the sum over $N_3$ in
\eq{Z} becomes divergent and the mean volume, $\mv{N_3}$, tends to the
infinity. In the vicinity of $\mu_c$ one could expect to find critical
behavior corresponding to a continuum limit of the model.

For this scenario, it is crucial that $\mu_c$ is finite:
$0<\mu_c<+\infty$, \ie the number of complexes as a function of the
volume, $N_3$, should grow at most exponentially \cite{ADJ}.
The purpose of the
present paper is to show that it is indeed the case. More precisely, we
prove \\
\noindent
{\sc theorem}: {\em There exists a finite constant $0<\mu^*<+\infty$
such that the number of 3-dimensional simply-connected simplicial
complexes, ${\cal N}(N_3)$, constructed of the given number of
3-simplexes, $N_3$, obeys the restriction}

\be
{\cal N}(N_3)<e^{\mu^*N_3}
\label{bound}
\ee
{\em for big enough values of $N_3$}.\\

``Simply-connected'' means that all loops in a complex $C$ are
contractible, \ie $\pi_1(C)=0$.
If the Poincar\'{e} hypothesis is true, this class of complexes
coincides with homeomorphic spheres. If not, it is somehow wider.
However, the statement of the theorem will obviously hold for the
spheres as well.

\section{Proof of the theorem}

We want to estimate ${\cal N}(N_3)$ from above being unable to calculate
it precisely. The general principle
can be formulated as follows. Let $S(\alpha)$ be a finite set of objects
defined by a set of descriptions $\alpha$. {\em If $\alpha\subset\beta$,
then $S(\beta)\subseteq S(\alpha)$ and $|S(\beta)|\leq|S(\alpha)|$},
where $|S|$ denotes the number of objects in a set $S$.
In other words, being less specific, one cannot decrease the number of
objects.

Throughout the paper we use the following trick: {\em if
$S=\bigcup_{\{A\}}B_A$, then $|S|\leq |A| {\rm max}_{\{A\}}|B_A|$} and,
{\em if all $B_A\neq\emptyset$, then $|A|\leq |S|$}.
Less formally, let us notice that the existence of the exponential bound
(\ref{bound}) is not spoiled, if one weights every complex with a non-zero
weight which itself depends on $N_3$  and grows at most exponentially.
Also, if one manages to represent a sum over complexes as a weighted
sum over another class of objects and to prove that the weights are
exponentially bounded, then one can put them all equal to 1 and simply
estimate the number of objects in the new class.

The simplest example of a weighted sum over complexes is given by the
micro-canonical partition function $Z_{N_3}(\alpha)$ in \eq{Z}. Another
example of physical interest is the partition function in the presence
of free matter fields. To introduce it let us consider the set of
cell complexes dual to simplicial ones. Their 1-skeletons are some
$\phi^4$ Feynman graphs whose vertices correspond to tetrahedra and
propagators are dual to triangles. Such a graph can be defined by the
adjacency matrix

\be
G_{ij}=\left\{\ba{ll} 1 & \mbox{if vertices $i$ and $j$ are connected by
a link}\\
0 & \mbox{otherwise} \ea \right.
\ee

Of course, the set of $\phi^4$ graphs is not identical to the one of
simplicial complexes, similarly as, in two dimensions, the set of ordinary
$\phi^3$ diagrams is different from the one of triangulations (one has
to introduce the notion of fat graphs to establish the equivalence).
However, we shall loosely use the term ``$\phi^4$-diagram'' because of
its commonness.

If one attaches the $n$-component free matter field $x^{\mu}_i\
(\mu=1,\ldots,n)$ to the $i$-th vertex, the resulting gaussian integral
can be performed explicitly for every given graph $G$:

\be
\int \prod_{i=1}^{N_3-1}\prod_{\mu=1}^n d x^{\mu}_i \exp\Big[ -\frac12
\sum_{i,j=1}^{N_3-1}G_{ij}(x^{\mu}_i-x^{\mu}_j)^2\Big] =
\det\mbox{}^{-\frac{n}2}L
\label{gint}
\ee
where $L$ is the discrete Laplacian

\be
L_{ij}=4\delta_{ij}-G_{ij}
\ee
for the determinant of which there is the nice combinatorial
representation given by the Kirchhoff theorem:

\be
\det L = |T(G)|
\ee
where $T(G)$ is the set of all maximal connected trees embedded into the
$\phi^4$ graph $G$ or, equivalently, the number of ways
to cut links of the graph such that it becomes a connected tree.
To kill the zero mode in \eq{gint}, we fixed the field at the $N_3$-th
vertex, therefore we consider rooted trees with the root attached to
this vertex.

The case we are interested in is the 2-component Grassmann field, where
we obtain the partition function\footnote{$\mbox{}^)$
This model was considered by
V.A.Kazakov several years ago as the simplest generalization of the
corresponding 2-dimensional matrix model \cite{privcom}.}

\be
{\cal Z}^{(1)}(\alpha,\mu)=\sum_{\{S^3\}}\det L\; e^{\alpha N_0 - \mu N_3}
=\sum_{N_3}^{\infty} Z^{(1)}_{N_3}(\alpha)e^{-\mu N_3}
\label{Z1}
\ee
The weights, $ \det L$, are non-zero integers, hence,

\be
Z_{N_3}(\alpha)<Z^{(1)}_{N_3}(\alpha)
\ee

In two dimensions, this type of matter has the central charge $c=-2$ and the
corresponding matrix model has been solved explicitly \cite{KKM}.
We can repeat the same trick in 3 dimensions:

\be
Z^{(1)}_{N_3}(\alpha)=\sum_{\{G\}}\det L\; e^{\alpha N_0}
=\sum_{\{G\}}\sum_{\{T(G)\}}e^{\alpha N_0}=\sum_{\{T\}}
\sum_{\{G(T)\}}e^{\alpha N_0}
\ee
Namely, we have the sum over complexes ($\phi^4$ graphs),
$\sum_{\{G\}}$, weighted with
the number of maximal trees in each $\sum_{\{T(G)\}}$. We can obtain
exactly the same configurations taking the sum over all possible
trees, $\sum_{\{T\}}$, weighted with the number of complexes,
$\sum_{\{G(T)\}}$, which can be recovered from a given tree, $T$, by
restoring cut links.

In 3 dimensions, the appearing trees are dual to spherical
simplicial balls obtained from a single tetrahedron
by subsequently gluing other tetrahedra to faces of the boundary.
If we denote $n_0,\ n_1,\ n_2$ the numbers of vertices, links and
triangles on the boundary of a ball, we find that

\be
n_0=N_3+3 \hspace{1pc} n_2=2(N_3+1)
\ee

The set of the boundaries are a subset of all planar
spherical triangulations, and
their number is exponentially bounded as a function of $N_3$.
Therefore, the remaining problem is to estimate the maximal number of closed
3-dimensional simplicial spheres which can be obtained from an
arbitrary spherical ball by pair-wise identifying triangles on its
boundary.

It seems to be rather  difficult to find a criterion which would single
out only homeomorphic spheres from all possible configurations.
Therefore, we demand instead that all final complexes
are to be simply-connected, \ie having the trivial fundamental group.

Let us consider loops embedded into
a $\phi^4$ diagram dual to a simplicial complex $C$ (which is the
1-skeleton of a dual cell complex, $\wt{C}$).
As any closed  loop is contractible, it can be represented as the
boundary of some 2-dimensional disc embedded into the complex $\wt{C}$.
It means simply that, to shrink a loop, one has
to pull it subsequently through 2-cells of $\wt{C}$
(they are in correspondence with links in $C$).
Obviously, we have to allow for self-intersections and multiple
coverings of the same face.  However, any $k$-cell of the disc has to
cover one and only one $k$-cell of $\wt{C}$. In other words, we
consider only regular (\ie locally unambiguous) embeddings.

If two triangles on the surface of a ball are identified, there is
always a closed in $\wt{C}$ loop represented by a segment lying completely
inside the ball with the ends attached to the identified triangles.
Having cut $\wt{C}$, one cuts any disc spanned by the loop. The discs
are represented by planar graphs and the cut links form what is called
``rainbow diagrams'', \ie collections of self-avoiding segments. If a cut
disc remains connected, these segments form rainbows consisting of
embedded arches.  The deepest segment connects a pair of triangles
having a common side on the surface of the ball, thus recovering a
2-cell of $\wt{C}$. If the disc is disconnected by cutting $\wt{C}$,
such configurations may be absent.
\bigskip

\noindent
{\sc Proposition~1} {\em Let $\wt{C}_{\rm b}$ be a cell complex dual
to a simplicial ball (\ie obtained from some closed complex $\wt{C}$
by cutting its links). If, by cutting a link in $\wt{C}_{\rm b}$, one
disconnects all discs embedded into $\wt{C}_{\rm b}$ and spanned by
loops going through the link, while the complex remains connected,
then $\wt{C}_{\rm b}$ is not simply connected.}

\noindent
{\sc Proof:} If a cut disconnects all the discs, it means that the
link enters at least twice in any closed contractible loop going
through it. As the complex remains connected, there must be another
path connecting the ends of the link. The union of this path and the
link represents a closed loop which cannot be a boundary of any disc,
hence, it is not contractible. \bigskip

Now, let us notice that, if one starts with an arbitrary closed
simply-connected complex, a final ball does not depend on a sequence
in which cuts are performed. One can choose a particular sequence
never creating non-contractible loops, owing to the following fact.

\bigskip
\noindent
{\sc Proposition~2} {\em Let $\wt{C}_{\rm b}$ be a cell complex
defined above. If, by cutting any link in $\wt{C}_{\rm b}$, one
creates a non-contractible loop, then $\wt{C}_{\rm b}$ is not
simply-connected.}

\noindent
{\sc Proof:} Let a cut create a new non-contractible loop. It is a
collection of links of $\wt{C}_{\rm b}$. Before the cut, it was
contractible, therefore there was a disc spanned by it. The cutting made
it a ring, hence, the cut link belongs to the disc. Moreover, it
belongs to any disc whose boundary is homotopic to the new loop. As it
holds for any link, we are considering an effectively 2-dimensional
object, \ie a collection of surfaces with holes having some identified
vertices, links and/or faces. If, by deleting a link, we create a
non-contractible loop in a simply-connected 2-dimensional space, it
means that it has the homotopy type of a disc. A disc must have a
boundary. However, we cannot  create a non-contractible loop by
cutting a link belonging to the boundary, Hence, we find a
contradiction: one cannot fulfill all the conditions simultaneously.
\bigskip

Thus,  any simply-connected
simplicial complex can be obtained from a ball by subsequently gluing up
triangles having a common side on the surface and such a pair always
exists. Obviously, the number of inequivalent sequences of such
identifications is
bigger than the actual number of complexes: two triangles may have two
common sides and it may happen at any step. It means an overcounting.
However, the gain is that we obtain a purely 2-dimensional problem now.

The boundary  of the ball can become a
collection of disconnected 2-dimensional spheres but any identification
of triangles belonging to different components produces a
non-contractible loop.

So, one starts with an arbitrary  2-dimensional
spherical $\phi^3$ diagram and applies the operation which can be
represented as the flip of a link with  the subsequent elimination of
it:
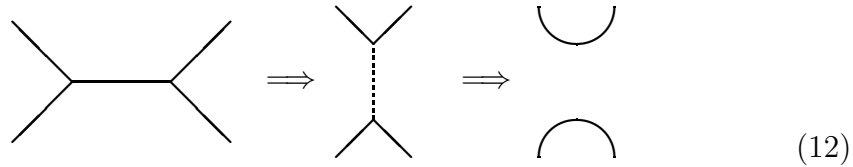
\setlength{\unitlength}{1mm}
\be
\bp(80,20)(-40,-10)
\thicklines
\put(-35,0){\line(-1,1){8}}
\put(-35,0){\line(-1,-1){8}}
\put(-35,0){\line(1,0){13}}
\put(-22,0){\line(1,-1){8}}
\put(-22,0){\line(1,1){8}}
\put(-11,-5){\makebox(10,10){$\Longrightarrow$}}
\put(5,5){\line(-1,1){5}}
\put(5,5){\line(1,1){5}}
\put(5,-5){\line(-1,-1){5}}
\put(5,-5){\line(1,-1){5}}
\multiput(5,-5)(0,1){10}{\line(0,1){0.5}}
\put(15,-5){\makebox(10,10){$\Longrightarrow$}}
\put(32,10){\oval(10,10)[b]}
\put(32,-10){\oval(10,10)[t]}
\ep
\ee
Both of these moves are well known and have been used in Monte-Carlo
simulations of triangulated surfaces \cite{KKM}. Having glued all triangles,
one finishes with a collection of
self-avoiding closed loops, the number of which equals $N_0-1$.

Let us notice that, instead of removing links, we can mark them with
dashed propagators. For it we need the infinite set of vertices

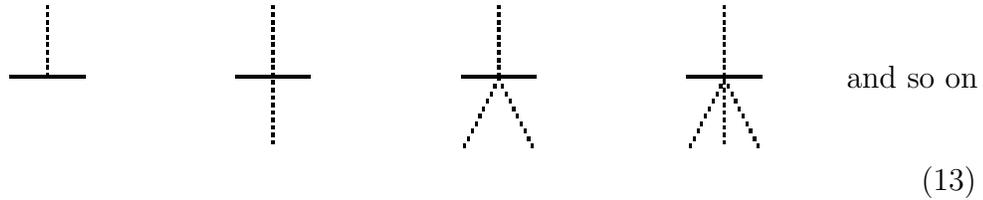
\be
\bp(130,20)(5,-10)
\thicklines
\multiput(10,0)(30,0){4}{\line(1,0){10}}
\multiput(15,0)(0,1){10}{\line(0,1){0.5}}
\multiput(45,0)(0,1){10}{\line(0,1){0.5}}
\multiput(75,0)(0,1){10}{\line(0,1){0.5}}
\multiput(105,0)(0,1){10}{\line(0,1){0.5}}
\multiput(45,0)(0,-1){10}{\line(0,1){0.5}}
\multiput(75,0)(-0.5,-1){10}{\line(0,-1){0.5}}
\multiput(75,0)(0.5,-1){10}{\line(0,-1){0.5}}
\multiput(105,0)(0,-1){10}{\line(0,1){0.5}}
\multiput(105,0)(-0.5,-1){10}{\line(0,-1){0.5}}
\multiput(105,0)(0.5,-1){10}{\line(0,-1){0.5}}
\put(125,-5){\makebox(10,10){and so on}}
\ep
\label{vert}
\ee

\noindent
which are generated by flips. For example,\\

\noindent
\bp(25,20)(-10,-10)
\thicklines
\put(-5,0){\line(1,0){10}}
\put(-5,0){\line(-1,-1){5}}
\put(-5,0){\line(-1,1){5}}
\put(5,0){\line(1,-1){5}}
\put(5,0){\line(1,1){5}}
\multiput(0,0)(0,-1){8}{\line(0,-1){0.5}}
\ep
\raisebox{9mm}{gives}
\bp(18,20)(-10,-10)
\thicklines
\put(-5,7){\line(1,0){10}}
\put(-5,0){\line(1,0){10}}
\multiput(0,0)(0,1){8}{\line(0,1){0.5}}
\multiput(0,0)(0,-1){8}{\line(0,-1){0.5}}
\ep
\raisebox{9mm}{\Large ;\hspace{5mm}}
\bp(25,20)(-10,-10)
\thicklines
\put(-5,0){\line(1,0){10}}
\put(-5,0){\line(-1,-1){5}}
\put(-5,0){\line(-1,1){5}}
\put(5,0){\line(1,-1){5}}
\put(5,0){\line(1,1){5}}
\multiput(-2,0)(0,-1){8}{\line(0,-1){0.5}}
\multiput(2,0)(0,-1){8}{\line(0,-1){0.5}}
\ep
\raisebox{9mm}{gives}
\bp(20,20)(-10,-10)
\thicklines
\put(-5,7){\line(1,0){10}}
\put(-5,0){\line(1,0){10}}
\multiput(0,0)(0,1){8}{\line(0,1){0.5}}
\multiput(0,0)(0.5,-1){8}{\line(0,-1){0.5}}
\multiput(0,0)(-0.5,-1){8}{\line(0,-1){0.5}}
\ep
\raisebox{9mm}{and so on}

\noindent
\bp(22,20)(-10,-10)
\thicklines
\put(-5,0){\line(1,0){10}}
\put(-5,0){\line(-1,-1){5}}
\put(-5,0){\line(-1,1){5}}
\put(5,0){\line(1,-1){5}}
\put(5,0){\line(1,1){5}}
\multiput(-2,0)(0,-1){8}{\line(0,-1){0.5}}
\multiput(2,0)(0,1){8}{\line(0,-1){0.5}}
\ep
\raisebox{9mm}{or}
\bp(22,20)(-10,-10)
\thicklines
\put(-5,0){\line(1,0){10}}
\put(-5,0){\line(-1,-1){5}}
\put(-5,0){\line(-1,1){5}}
\put(5,0){\line(1,-1){5}}
\put(5,0){\line(1,1){5}}
\multiput(-2,0)(0,1){8}{\line(0,-1){0.5}}
\multiput(2,0)(0,-1){8}{\line(0,-1){0.5}}
\ep
\raisebox{9mm}{or}
\bp(22,20)(-10,-10)
\thicklines
\put(-5,0){\line(1,0){10}}
\put(-5,0){\line(-1,-1){5}}
\put(-5,0){\line(-1,1){5}}
\put(5,0){\line(1,-1){5}}
\put(5,0){\line(1,1){5}}
\multiput(0,0)(0,1){8}{\line(0,-1){0.5}}
\multiput(0,0)(0,-1){8}{\line(0,-1){0.5}}
\ep
\raisebox{9mm}{give}
\bp(20,20)(-10,-10)
\thicklines
\put(-5,3){\line(1,0){10}}
\put(-5,-3){\line(1,0){10}}
\multiput(0,-10)(0,1){20}{\line(0,1){0.5}}
\ep

\noindent
\bp(22,20)(-10,-10)
\thicklines
\put(-5,0){\line(1,0){10}}
\put(-5,0){\line(-1,-1){5}}
\put(-5,0){\line(-1,1){5}}
\put(5,0){\line(1,-1){5}}
\put(5,0){\line(1,1){5}}
\multiput(-3,0)(0,-1){8}{\line(0,-1){0.5}}
\multiput(0,0)(0,-1){8}{\line(0,-1){0.5}}
\multiput(3,0)(0,1){8}{\line(0,-1){0.5}}
\ep
\raisebox{9mm}{or}
\bp(22,20)(-10,-10)
\thicklines
\put(-5,0){\line(1,0){10}}
\put(-5,0){\line(-1,-1){5}}
\put(-5,0){\line(-1,1){5}}
\put(5,0){\line(1,-1){5}}
\put(5,0){\line(1,1){5}}
\multiput(-3,0)(0,-1){8}{\line(0,-1){0.5}}
\multiput(0,0)(0,1){8}{\line(0,-1){0.5}}
\multiput(3,0)(0,-1){8}{\line(0,-1){0.5}}
\ep
\raisebox{9mm}{or}
\bp(22,20)(-10,-10)
\thicklines
\put(-5,0){\line(1,0){10}}
\put(-5,0){\line(-1,-1){5}}
\put(-5,0){\line(-1,1){5}}
\put(5,0){\line(1,-1){5}}
\put(5,0){\line(1,1){5}}
\multiput(-3,0)(0,1){8}{\line(0,-1){0.5}}
\multiput(0,0)(0,-1){8}{\line(0,-1){0.5}}
\multiput(3,0)(0,-1){8}{\line(0,-1){0.5}}
\ep
\raisebox{9mm}{or}
\bp(22,20)(-10,-10)
\thicklines
\put(-5,0){\line(1,0){10}}
\put(-5,0){\line(-1,-1){5}}
\put(-5,0){\line(-1,1){5}}
\put(5,0){\line(1,-1){5}}
\put(5,0){\line(1,1){5}}
\multiput(0,0)(-0.5,-1){8}{\line(0,-1){0.5}}
\multiput(0,0)(0.5,-1){8}{\line(0,-1){0.5}}
\multiput(0,0)(0,1){8}{\line(0,-1){0.5}}
\ep
\raisebox{9mm}{give}
\bp(20,20)(-10,-8)
\thicklines
\put(-5,6){\line(1,0){10}}
\put(-5,0){\line(1,0){10}}
\multiput(0,0)(0,1){12}{\line(0,1){0.5}}
\multiput(0,0)(0.5,-1){8}{\line(0,-1){0.5}}
\multiput(0,0)(-0.5,-1){8}{\line(0,-1){0.5}}
\ep

\noindent
and so on.

In the end,
we obtain planar diagrams with two types of propagators: solid ones
produce closed loops while dashed form arbitrary clusters. To recover
initial diagrams one has to flip dashed links in all possible ways. For
any given cluster, the number of final configurations is a subset of all
planar $\phi^3$ diagrams, hence, is exponentially bounded as a function
of the number of links. Let us show it. At first, flips  have to be
performed in linear clusters which gives all possible planar $\phi^3$
trees doubled in a ``mirror'':

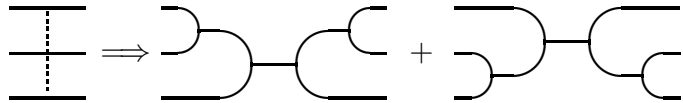
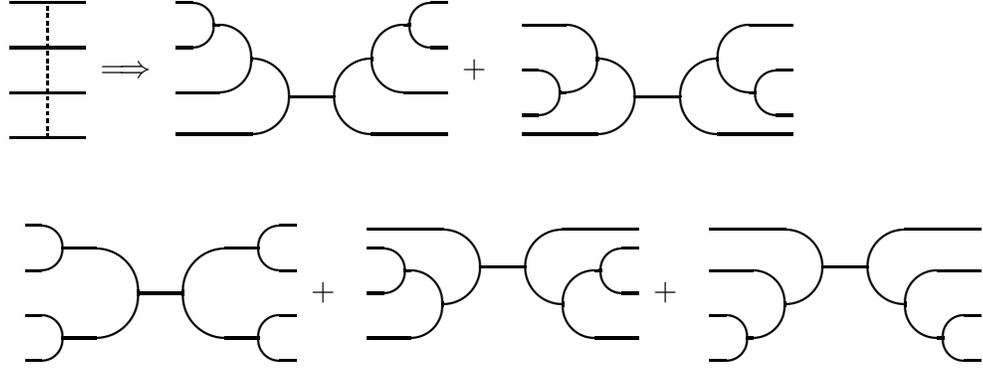
\bp(12,20)(-5,-10)
\thicklines
\put(-5,6){\line(1,0){10}}
\put(-5,0){\line(1,0){10}}
\put(-5,-6){\line(1,0){10}}
\multiput(0,0)(0,1){6}{\line(0,1){0.5}}
\multiput(0,0)(0,-1){6}{\line(0,1){0.5}}
\ep
\raisebox{9mm}{$\Longrightarrow$}
\bp(33,20)(-15,-10)
\thicklines
\put(-3,-1.5){\line(1,0){6}}
\put(10,-1.5){\oval(14,9)[l]}
\put(-10,-1.5){\oval(14,9)[r]}
\put(15,3){\oval(10,6)[l]}
\put(-15,3){\oval(10,6)[r]}
\put(-15,-6){\line(1,0){6}}
\put(15,-6){\line(-1,0){6}}
\ep
\raisebox{9mm}{+\ }
\bp(33,20)(-15,-10)
\thicklines
\put(-3,1.5){\line(1,0){6}}
\put(10,1.5){\oval(14,9)[l]}
\put(-10,1.5){\oval(14,9)[r]}
\put(15,-3){\oval(10,6)[l]}
\put(-15,-3){\oval(10,6)[r]}
\put(-15,6){\line(1,0){6}}
\put(15,6){\line(-1,0){6}}
\ep

\bp(12,20)(-5,-10)
\thicklines
\put(-5,9){\line(1,0){10}}
\put(-5,3){\line(1,0){10}}
\put(-5,-3){\line(1,0){10}}
\put(-5,-9){\line(1,0){10}}
\multiput(0,-9)(0,1){18}{\line(0,1){0.5}}
\ep
\raisebox{9mm}{$\Longrightarrow$}
\bp(40,20)(-20,-10)
\thicklines
\put(-3,-3.5){\line(1,0){6}}
\put(8,-3.5){\oval(10,10)[l]}
\put(-8,-3.5){\oval(10,10)[r]}
\put(8,-8.5){\line(1,0){10}}
\put(-8,-8.5){\line(-1,0){10}}
\put(13,1.5){\oval(10,9)[l]}
\put(-13,1.5){\oval(10,9)[r]}
\put(13,-3){\line(1,0){5}}
\put(-13,-3){\line(-1,0){5}}
\put(18,6){\oval(10,6)[l]}
\put(-18,6){\oval(10,6)[r]}
\ep
\raisebox{9mm}{+\ }
\bp(40,20)(-20,-10)
\thicklines
\put(-3,-3.5){\line(1,0){6}}
\put(8,-3.5){\oval(10,10)[l]}
\put(-8,-3.5){\oval(10,10)[r]}
\put(8,-8.5){\line(1,0){10}}
\put(-8,-8.5){\line(-1,0){10}}
\put(13,1.5){\oval(10,9)[l]}
\put(-13,1.5){\oval(10,9)[r]}
\put(13,6){\line(1,0){5}}
\put(-13,6){\line(-1,0){5}}
\put(18,-3){\oval(10,6)[l]}
\put(-18,-3){\oval(10,6)[r]}
\ep
\eol

\bp(40,20)(-20,-10)
\thicklines
\put(-3,0){\line(1,0){6}}
\put(13,0){\oval(20,12)[l]}
\put(-13,0){\oval(20,12)[r]}
\put(18,6){\oval(10,6)[l]}
\put(-18,6){\oval(10,6)[r]}
\put(18,-6){\oval(10,6)[l]}
\put(-18,-6){\oval(10,6)[r]}
\ep
\raisebox{9mm}{+\ }
\bp(40,20)(-20,-10)
\thicklines
\put(-3,3.5){\line(1,0){6}}
\put(8,3.5){\oval(10,10)[l]}
\put(-8,3.5){\oval(10,10)[r]}
\put(8,8.5){\line(1,0){10}}
\put(-8,8.5){\line(-1,0){10}}
\put(13,-1.5){\oval(10,9)[l]}
\put(-13,-1.5){\oval(10,9)[r]}
\put(13,-6){\line(1,0){5}}
\put(-13,-6){\line(-1,0){5}}
\put(18,3){\oval(10,6)[l]}
\put(-18,3){\oval(10,6)[r]}
\ep
\raisebox{9mm}{+\ }
\bp(40,20)(-20,-10)
\thicklines
\put(-3,3.5){\line(1,0){6}}
\put(8,3.5){\oval(10,10)[l]}
\put(-8,3.5){\oval(10,10)[r]}
\put(8,8.5){\line(1,0){10}}
\put(-8,8.5){\line(-1,0){10}}
\put(13,-1.5){\oval(10,9)[l]}
\put(-13,-1.5){\oval(10,9)[r]}
\put(13,3){\line(1,0){5}}
\put(-13,3){\line(-1,0){5}}
\put(18,-6){\oval(10,6)[l]}
\put(-18,-6){\oval(10,6)[r]}
\ep

\noindent
and so on. The number of configurations with $n$ dashed links, $C_n$, is
given by Catalan's numbers generated by the function

\be
C(x)=\sum_{n=0}^{\infty}C_nx^n:=\sum_{n=0}^{\infty}\
\raisebox{-1cm}{$
\overbrace{
\bp(10,20)(-5,-10)
\thicklines
\put(0,0){\circle{10}}
\put(-1.5,-1.5){\mbox{\sf\bf T}}
\put(0,5){\line(0,1){5}}
\put(1,5){\line(1,3){1.66}}
\put(2,4.5){\line(2,3){3.33}}
\put(-1,5){\line(-1,3){1.66}}
\put(-2,4.5){\line(-2,3){3.33}}
\put(0,-5){\line(0,-1){5}}
\ep
}^{n+1}$}\
x^n=\frac1{2x}(1-\sqrt{1-4x})
\ee

After that dashed propagators attached to the ends of the
linear cluster have to be
reattached in all possible ways to links of these ``doubled trees''. The
number of configurations can be estimated as follows.

First, we have to calculate the generating function for the number of
possible attachments of dashed propagators to a solid line:

\be
q(\nu,\mu):=\sum_{n,m}\
\raisebox{-1cm}{$
\overbrace{\underbrace{
\bp(20,20)(-10,-10)
\thicklines
\put(0,0){\circle*{3}}
\put(-8,0){\line(1,0){16}}
\multiput(0,0)(0,1){10}{\line(0,1){0.5}}
\multiput(0,0)(0.5,1){9}{\line(0,1){0.5}}
\multiput(0,0)(0.75,0.75){9}{\line(0,1){0.5}}
\multiput(0,0)(-0.5,1){9}{\line(0,1){0.5}}
\multiput(0,0)(-0.75,0.75){9}{\line(0,1){0.5}}
\multiput(0,0)(0,-1){10}{\line(0,-1){0.5}}
\multiput(0,0)(-0.5,-1){9}{\line(0,-1){0.5}}
\multiput(0,0)(-0.75,-0.75){9}{\line(0,-1){0.5}}
\multiput(0,0)(0.5,-1){9}{\line(0,-1){0.5}}
\multiput(0,0)(0.75,-0.75){9}{\line(0,-1){0.5}}
\ep
}_m
}^{n}$}\
\nu^n\mu^m
\ee
It can be obtained from the equation

\be
q(\nu,\mu)=\Big[\frac{\nu}{1-\mu}+\frac{\mu}{1-\nu}-\nu\mu\Big]
q(\nu,\mu)+1
\ee
which gives

\be
q(\nu,\mu)=\frac{(1-\nu)(1-\mu)}{(1-\nu-\mu)^2-(\nu+\mu)\nu\mu
+\nu^2\mu^2}
\ee

Second, we have to know the generating function for the number of all
possible attachments of dashed propagators to a $k$-legged tree on the
left (or on the right but not on both sides):

\be
t(x,\nu):=\sum_{k,n} \raisebox{-1mm}{${\scr n}\Big\{$}\hspace{5mm}
\raisebox{-1cm}{$
\overbrace{
\bp(10,20)(-5,-10)
\thicklines
\put(0,0){\circle{10}}
\put(-1.5,-1.5){\mbox{\sf\bf T}}
\put(0,5){\line(0,1){5}}
\put(1,5){\line(1,3){1.66}}
\put(2,4.5){\line(2,3){3.33}}
\put(-1,5){\line(-1,3){1.66}}
\put(-2,4.5){\line(-2,3){3.33}}
\put(0,-5){\line(0,-1){5}}
\multiput(-5,0)(-1,0){5}{\line(-1,0){0.5}}
\multiput(-5,0.5)(-1,0.25){5}{\line(-1,0){0.5}}
\multiput(-5,1)(-1,0.5){5}{\line(-1,0){0.5}}
\multiput(-5,-0.5)(-1,-0.25){5}{\line(-1,0){0.5}}
\multiput(-5,-1)(-1,-0.5){5}{\line(-1,0){0.5}}
\ep
}^{k+1}$}\
x^{k}\nu^n;
\hspace{2pc} t(x,0)=C(x)
\ee
The equation for it is simply

\be
t(x,\nu)=xC(x)\frac{t(x,\nu)}{1-\nu}+1
\ee
which gives

\be
t(x,\nu)=\frac{1-\nu}{1-\nu-xC(x)}
\ee
Then, the quantity we need to calculate is

\[
T(x,\nu,\mu)=\sum_{n,m,k}
\raisebox{-1mm}{${\scr n}\Big\{$}\hspace{5mm}
\raisebox{-5mm}{$
\underbrace{
\bp(20,10)(-10,-5)
\thicklines
\multiput(-10,0)(1,0){20}{\line(1,0){0.5}}
\multiput(-10,-5)(5,0){5}{\line(0,1){10}}
\multiput(-10,0)(-1,0){5}{\line(-1,0){0.5}}
\multiput(-10,0.5)(-1,0.25){5}{\line(-1,0){0.5}}
\multiput(-10,1)(-1,0.5){5}{\line(-1,0){0.5}}
\multiput(-10,-0.5)(-1,-0.25){5}{\line(-1,0){0.5}}
\multiput(-10,-1)(-1,-0.5){5}{\line(-1,0){0.5}}
\multiput(10,0)(1,0){5}{\line(1,0){0.5}}
\multiput(10,0.5)(1,0.25){5}{\line(1,0){0.5}}
\multiput(10,1)(1,0.5){5}{\line(1,0){0.5}}
\multiput(10,-0.5)(1,-0.25){5}{\line(1,0){0.5}}
\multiput(10,-1)(1,-0.5){5}{\line(1,0){0.5}}
\ep
}_{k+1}$}\hspace{5mm}
\raisebox{-1mm}{$\Big\}{\scr m}$}
\; x^{k}\nu^n\mu^m
=\sum_{n,m,k}\raisebox{-3cm}{
\bp(30,60)(-15,-30)
\thicklines
\put(0,-5){\line(0,1){10}}
\put(0,0){\circle*{3}}
\multiput(0,0)(-1,0){8}{\line(-1,0){0.5}}
\multiput(0,0.5)(-1,0.25){7}{\line(-1,0){0.5}}
\multiput(0,1)(-1,0.5){6}{\line(-1,0){0.5}}
\multiput(0,-0.5)(-1,-0.25){7}{\line(-1,0){0.5}}
\multiput(0,-1)(-1,-0.5){6}{\line(-1,0){0.5}}
\multiput(0,0)(1,0){8}{\line(1,0){0.5}}
\multiput(0,0.5)(1,0.25){7}{\line(1,0){0.5}}
\multiput(0,1)(1,0.5){6}{\line(1,0){0.5}}
\multiput(0,-0.5)(1,-0.25){7}{\line(1,0){0.5}}
\multiput(0,-1)(1,-0.5){6}{\line(1,0){0.5}}
\put(0,15){\oval(20,20)[b]}
\put(-10,20){\circle{10}}
\put(-11.5,18.5){\mbox{\sf\bf T$\mbox{}_1$}}
\put(-10,25){\line(0,1){5}}
\put(-9,25){\line(1,3){1.66}}
\put(-8,24.5){\line(2,3){3.33}}
\put(-11,25){\line(-1,3){1.66}}
\put(-12,24.5){\line(-2,3){3.33}}
\multiput(-15,20)(-1,0){5}{\line(-1,0){0.5}}
\multiput(-15,20.5)(-1,0.25){5}{\line(-1,0){0.5}}
\multiput(-15,21)(-1,0.5){5}{\line(-1,0){0.5}}
\multiput(-15,19.5)(-1,-0.25){5}{\line(-1,0){0.5}}
\multiput(-15,19)(-1,-0.5){5}{\line(-1,0){0.5}}
\put(10,20){\circle{10}}
\put(8.5,18.5){\mbox{\sf\bf T$\mbox{}_2$}}
\put(10,25){\line(0,1){5}}
\put(9,25){\line(-1,3){1.66}}
\put(8,24.5){\line(-2,3){3.33}}
\put(11,25){\line(1,3){1.66}}
\put(12,24.5){\line(2,3){3.33}}
\multiput(15,20)(1,0){5}{\line(1,0){0.5}}
\multiput(15,20.5)(1,0.25){5}{\line(1,0){0.5}}
\multiput(15,21)(1,0.5){5}{\line(1,0){0.5}}
\multiput(15,19.5)(1,-0.25){5}{\line(1,0){0.5}}
\multiput(15,19)(1,-0.5){5}{\line(1,0){0.5}}
\put(0,-15){\oval(20,20)[t]}
\put(-10,-20){\circle{10}}
\put(-11.5,-21.5){\mbox{\sf\bf T$\mbox{}_1$}}
\put(-10,-25){\line(0,-1){5}}
\put(-9,-25){\line(1,-3){1.66}}
\put(-8,-24.5){\line(2,-3){3.33}}
\put(-11,-25){\line(-1,-3){1.66}}
\put(-12,-24.5){\line(-2,-3){3.33}}
\multiput(-15,-20)(-1,0){5}{\line(-1,0){0.5}}
\multiput(-15,-20.5)(-1,-0.25){5}{\line(-1,0){0.5}}
\multiput(-15,-21)(-1,-0.5){5}{\line(-1,0){0.5}}
\multiput(-15,-19.5)(-1,0.25){5}{\line(-1,0){0.5}}
\multiput(-15,-19)(-1,0.5){5}{\line(-1,0){0.5}}
\put(10,-20){\circle{10}}
\put(8.5,-21.5){\mbox{\sf\bf T$\mbox{}_2$}}
\put(10,-25){\line(0,-1){5}}
\put(9,-25){\line(-1,-3){1.66}}
\put(8,-24.5){\line(-2,-3){3.33}}
\put(11,-25){\line(1,-3){1.66}}
\put(12,-24.5){\line(2,-3){3.33}}
\multiput(15,-20)(1,0){5}{\line(1,0){0.5}}
\multiput(15,-20.5)(1,-0.25){5}{\line(1,0){0.5}}
\multiput(15,-21)(1,-0.5){5}{\line(1,0){0.5}}
\multiput(15,-19.5)(1,0.25){5}{\line(1,0){0.5}}
\multiput(15,-19)(1,0.5){5}{\line(1,0){0.5}}
\ep}
x^{k}\nu^n\mu^m
\]
\be
=q(\nu,\mu)\Bigg\{1+\frac{x(1-\nu)^2(1-\mu)^2}
{\Big((1-\nu)^2-xC(x)\Big)\Big((1-\mu)^2-xC(x)\Big)}\Bigg\}
\label{T}
\ee

Newly appearing clusters have to be expanded in all possible ways
according to the same procedure. What is important is that their number
is restricted by the coefficients of $q(\nu,\mu)$, which is a factor in
(\ref{T}), and does not depend on a volume of the trees. Therefore, one
simply iterates till all dashed links disappear.
The combinatorial reason for the existence of
an exponential bound  for the number of obtained configurations is
clear: they are all planar!

To calculate the number of ``unflipped'' configurations, we can use the
following matrix integral generating all vertices from (\ref{vert}) with
equal weights

\be\ba{c}
I={\displaystyle\int \prod_{a=1}^{\nu}}d^{\scs N^2}X_a\: d^{\scs N^2}Y
\exp\Big[-\frac{N}2{\displaystyle\sum_{a=1}^{\nu}}
\tr X^2_a -\frac{N}2\tr Y^2 + \mu N {\displaystyle\sum_{a=1}^{\nu}}
\tr \Big(YX_a\frac1{1-\mu Y}X_a\Big)\Big] \\[2pc]
={\displaystyle\int} d^{\scs N^2}Y
\exp\Big[-\frac{N}2\tr Y^2 -\frac{\nu}2\tr\mbox{}^2\log\Big(1\otimes 1
- \mu Y\otimes\frac1{1-\mu Y}- \frac1{1-\mu Y}\otimes \mu Y \Big) \Big]
\ea
\label{I}
\ee
which reminds the Kostov's representation for the $O(n)$ matrix
model in the dense phase \cite{O(n)}.
We have attached the lower index to the gaussian $X$ variable
to weight the closed loops. Thus,

\be
I(\nu,\mu)=\sum_{\{{\cal D}\}} \nu^{N_0-1}\mu^{N_3}
\ee
where $\sum_{\{{\cal D}\}}$ is the sum over all the diagrams. This matrix model
belongs to the same universality class as the $O(n)$-model in the
dense phase. $I(\nu,\mu)$ has a finite radius of convergence as a
function of $\mu$. Thus, the theorem is established.

\section{Discussion}

It is very difficult to take account of the entropy of configurations in
order to improve the representation (\ref{I}). Therefore, let us
consider the simplest self-consistent model, which is equivalent to the dense
phase of the $O(n)$ matrix model \cite{O(n)}:

\be
I_{\scs O(n)} =\int d^{\scs N^2}Y
\exp\Big[-\frac{N}2\tr Y^2 -\frac{\nu}2\tr\mbox{}^2\log\Big(1\otimes 1
- \mu Y\otimes 1 -  1 \otimes \mu Y \Big) \Big]
\label{I*}
\ee

Let us suppose that this truncated model
bears nevertheless some qualitative features of
3-dimensional simplicial gravity.

The critical behavior of the $O(n)$ matrix
model is well known \cite{O(n)}.
For $\nu<2$, it describes 2-d gravity interacting with $c<1$
conformal matter

\be
c=1-6\frac{(1-g_0)^2}{g_0}; \hspace{2pc} \nu=-2\cos \pi g_0
\ee
while for $\nu>2$ the
corresponding matter is non-critical and the model trivializes.
In this phase, the number of closed loops is proportional to the volume
and the mean length of each remains finite.

In the vicinity of a critical point $\mu_c$, the partition function
behaves as

\be
I_{\scs O(n)} \approx (\mu_c-\mu)^{2-\gamma_{\rm str}}
\ee
Two main quantities of interest are

\be
\mv{N_3}_{\mbox{}_{O(n)}}
=\mu\frac{\d\ }{\d\mu} \log I_{\scs O(n)} \approx -(2-\gamma_{\rm str})
\frac{\mu_c}{\mu_c-\mu}
\ee
and
\be\ba{rl}
\mv{N_0}_{\mbox{}_{O(n)}}
&=\nu\frac{\d\ }{\d\nu} \log I_{\scs O(n)} \approx (2-\gamma_{\rm str})
\frac{\nu\mu'_c(\nu)}{\mu_c(\nu)-\mu} - \nu\frac{\d\gamma_{\rm
str}}{\d\nu}\log|\mu_c-\mu|\\[2pc]
&\approx-\nu\frac{\mu'_c}{\mu_c}\mv{N_3}_{\mbox{}_{O(n)}}
+\nu\frac{\d\gamma_{\rm str}}{\d\nu}\log\mv{N_3}_{\mbox{}_{O(n)}}
\ea
\ee
where the coefficients can be calculated explicitly using the Gaudin and
Kostov's exact solution \cite{O(n)}

\be
\frac{\d\ }{\d\nu}\log\mu_c= -\frac12\frac1{\nu+2};
\hspace{2pc}
\frac{\d\gamma_{\rm str}}{\d\nu}=\frac1{2\pi g_0^2 \sin\pi(1-g_0)}
\ee
If $2-\nu\ll 1$, then

\be
\mv{N_0}_{\mbox{}_{O(n)}} \approx \frac14\mv{N_3}_{\mbox{}_{O(n)}}
+\frac1{\pi\sqrt{2-\nu}}\log\mv{N_3}_{\mbox{}_{O(n)}}
\ee

If $\nu>2$, then $\gamma_{\rm str}=-\frac12$ is independent of $\nu$ and
$\mv{N_0}_{\mbox{}_{O(n)}}$ is proportional to $\mv{N_3}_{\mbox{}_{O(n)}}$.

This behavior is strikingly similar to the results of numerical
simulations! In  \rf{3dsim2}, a phase transition in the model (\ref{Z})
with respect to the
fugacity for the number of vertices, $\alpha$, was found. In the
``cold'' phase, $\alpha>\alpha_c$, the mean number of vertices,
$\mv{N_0}$, is strictly proportional to a volume, $\mv{N_3}$. In the
``hot'' phase, $\alpha<\alpha_c$, $\mv{N_0}$ is a non-trivial function
of $\mv{N_3}$. The analogy with the $O(n)$ matrix model suggests that
the most probable scaling is\footnote{$\mbox{}^)$
Unfortunately, numerical
experiments cannot provide an accuracy sufficient to completely rule out
other possibilities, \eg a power scaling.} $\mv{N_0}\approx c_1\mv{N_3}
+c_2\log\mv{N_3}$ with $c_2$ singular at the critical point $\alpha_c$.
Presumably, it is a type of this singularity that can be, in principle,
calculated in continuum theory in order to compare predictions of
both approaches.
The obvious problem for such a hypothetical comparison is that
$\alpha$ is a bare coupling. Therefore, in the lattice
model, only critical
points with respect to it could show some universal features.

Of course, the $O(n)$ matrix model cannot give us precise information
about simplicial gravity. However, it is very plausible that it has
qualitatively the same phase structure as models (\ref{Z}) and (\ref{Z1})
and may be quite instructive from this point of view.

To conclude, let us express the hope that, when 3-dimensional
simplicial gravity is solved, it will reveal as much beautiful
mathematical and physical structure as the matrix models have been
doing.\\

\bigskip
{\Large\bf Acknowledgments}
\bigskip

This work has appeared as a result of the numerous discussions
with J.Ambj\o rn, B.Durhuus, T.Jonsson, V.A.Kazakov and A.A.Migdal,
who have for many years been sharing with me their ideas and knowledge.

Financial support from the EEC Program ``Human Capital and Mobility''
under the contract ERBCHBICT9941621 is greatfully acknowledged.

\end{document}